\documentclass[a4paper,11pt]{article} 
\usepackage[margin=2.5cm]{geometry}\usepackage{graphicx}
\usepackage{multirow}
\usepackage{tabularx} 
\usepackage{rotating,graphicx} 
\usepackage{appendix} 
\usepackage{makecell}
\usepackage{amsmath}
\usepackage{amssymb}
\usepackage{cite}
\usepackage[T1]{fontenc}
\usepackage{amsmath}
\usepackage[amssymb]{SIunits}
\usepackage{hyperref}

\usepackage{lineno}

\newcommand{\btwo}{\ensuremath{B_{2}}}

\newcommand{\tritium}{\ensuremath{{}^{3}\mathrm{H}}}
\newcommand{\hethree}{\ensuremath{{}^{3}\mathrm{He}}}
\newcommand{\hefour}{\ensuremath{{}^{4}\mathrm{He}}}
\newcommand{\hthreelambda}{\ensuremath{{}^{3}_{\Lambda}\mathrm{H}}}
\newcommand{\hfourlambda}{\ensuremath{{}^{4}_{\Lambda}\mathrm{H}}}
\newcommand{\hfourtwolambda}{\ensuremath{{}^{4}_{\Lambda\Lambda}\mathrm{H}}}
\newcommand{\hefourlambda}{\ensuremath{{}^{4}_{\Lambda}\mathrm{He}}}

\newcommand{\antihethree}{\ensuremath{{}^{3}\overline{\mathrm{He}}}}

\newcommand{\rmsradius}{\ensuremath{\lambda_{A}}}

\newcommand{\GeVc}{GeV/$c$}

\newcommand{\PbPb}         {\mbox{Pb--Pb}}
\newcommand{\pPb}          {\mbox{p--Pb}}

\newcommand{\avdNdeta}       {\ensuremath{\left<\mathrm{d}N_\mathrm{ch}/\mathrm{d}\eta_{lab}\right>}}

\newcommand{\pt}           {\ensuremath{p_{\mathrm{T}}}}

\begin{document}

\begin{center}

{\Large \textbf{Testing production scenarios for (anti-)(hyper-)nuclei with multiplicity-dependent measurements at the LHC}}

\medskip

Francesca Bellini\footnote{Presented at the XXV Cracow EPIPHANY Conference on Advances in Heavy Ion Physics}\footnote{Contacts: fbellini@cern.ch, akalweit@cern.ch} and Alexander P. Kalweit
\\European Organization for Nuclear Research (CERN), Geneva, Switzerland

\medskip

\end{center}

\begin{abstract}
The production of light anti- and hyper-nuclei provides unique observables to characterise the system created in high energy proton-proton (pp), proton-nucleus (pA) and nucleus-nucleus (AA) collisions. In particular, nuclei and hyper-nuclei are special objects with respect to non-composite hadrons (such as pions, kaons, protons, etc.), because their size is comparable to a fraction or the whole system created in the collision. Their formation is typically described within the framework of coalescence and thermal-statistical production models. 
In order to distinguish between the two production scenarios, we propose to measure the coalescence parameter $B_A$ for different anti- and hyper-nuclei (that differ by mass, size and internal wave-function) as a function of the size of the particle emitting source. 
The latter can be controlled by performing systematic measurements of light anti- hyper- nuclei in different collision systems (pp, pA, AA) and as a function of the multiplicity of particles created in the collision. 
While it is often argued that the coalescence and the thermal model approach give very similar predictions for the production of light nuclei in heavy-ion collisions, our study shows that large differences can be expected for hyper-nuclei with extended wave-functions, as the hyper-triton. We compare the model predictions with data from the ALICE experiment and we discuss perspectives for future measurements with the upgraded detectors during the High-Luminosity LHC phase in the next decade. 
\end{abstract}

  
\section{Introduction: the ``anti-nuclei'' puzzle}
The formation of light anti- and hyper-nuclei in high energy proton-proton (pp), proton-nucleus (pA) and nucleus-nucleus (AA) collisions provides unique observables for the study of the system created in these reactions, and can be used to understand both the internal structure and the formation mechanisms of loosely-bound composite objects. 
The production of (anti-)(hyper-)nuclei in high-energy collisions is commonly described by following two distinct approaches: formation by nucleon coalescence at the system (kinetic) freeze-out~\cite{Butler:1963, Kapusta:1980, Scheibl:1998tk, Blum:2017qnn} or thermal-statistical production at the chemical freeze-out~\cite{Andronic:2010qu,Andronic:2017}.
Thanks to the large data samples of pp, \pPb~and \PbPb~collisions collected during the first ten years of operations of the CERN Large Hadron Collider (LHC), A Large Ion Collider Experiment (ALICE) Collaboration 
has measured the production of light nuclei and anti-nuclei at several centre-of-mass energies~\cite{ALICE:deuteronppPbPb2015, Adam:2015yta, ALICE:nucleipp2017, Acharya:2017dmc, Puccio:2019oyd, Acharya:2019rgc}, thus providing a crucial experimental input and a boost to theoretical and phenomenological investigations~\cite{Mrowczynski:2016xqm, Bellini:2018epz, Bazak:2018hgl, Zhao:2018lyf, Sun:2018mqq, Xu:2018jff, Oliinychenko:2018ugs}. 
In small collision systems, the experimental results seem to confirm the validity of the coalescence picture, with the most recent multiplicity-differential measurements pointing toward a dependence of the coalescence process on the volume of the particle-emitting source (''source size'' hereafter). 
In heavy-ion collisions, coalescence approaches that do not take into account the source size are not able to reproduce the data. 
At the same time, the production of light nuclei, anti-nuclei and hypertriton as measured in \PbPb~collisions is found to be consistent with statistical-thermal model predictions and a non-zero deuteron elliptic flow is observed.
This is surprising as (anti-)nuclei produced at chemical freeze-out are not expected to survive the hadronic phase: the deuteron is a ``fragile object'' when surrounded by the fireball created in heavy-ion collisions, because its binding energy ($B_E = 2.2$ MeV) is much lower than the characteristic temperatures of the system ($T_{chem} \approx 153$ MeV, $T_{kin} \approx 100$ MeV). Moreover, the cross-section for pion-induced deuteron breakup is significantly larger than the typical (pseudo)-elastic cross-sections for the re-scattering of hadronic resonance decay products \cite{Garcilazo:1982yc, Bass:1998ca, Schukraft:2017nbn, Oliinychenko:2018ugs}. 
These observations pose the ``(anti-)nuclei puzzle'': how can loosely-bound composite objects survive in the 
dense and hot fireball, freeze-out and develop collective flow like the other light-flavour non-composite hadrons? 

In our study~\cite{Bellini:2018epz} we have extended and combined known formalisms used to describe (anti-)(hyper-)nuclei production in order to allow, for the first time, a direct comparison of the thermal and coalescence models as well as a direct comparison to the ALICE data. 
Identifying the coalescence parameter ($B_A$) as the key observable, we present a consistent picture across different collision systems (pp, \pPb, \PbPb) for light (anti-)(hyper-)nuclei with mass number $A = 2, 3$ and 4.
We also suggest to address the open questions by looking at the production of nuclei and hyper-nuclei up to $A = 4$ that differ by size and properties, measured as a function of multiplicity used as a proxy for the source size. 
Whereas the (anti-)deuteron production has been measured multi-differentially and quite precisely with the LHC Run 1 and 2 data, the study of heavier objects with $A=3$ and 4 will greatly profit from the increase in integrated luminosity foreseen at the LHC Runs 3 and 4 in all collision systems~\cite{Citron:2018lsq}. 
A comprehension of (anti-)nuclei production mechanisms is not only relevant for nuclear and hadronic physics, but has applications in astrophysics and indirect Dark Matter searches~\cite{Aramaki:2015pii}. In recent years, it has been suggested that the detection of light anti-nuclei in space could provide a signature for the presence of Dark Matter in the Cosmos, see for instance~\cite{Cirelli:2014qia, Korsmeier:2017xzj}. 
Anti-deuterons and \antihethree~might indeed be produced by coalescence of antiprotons and antineutrons coming from the annihilation of Weakly Interacting Massive Particles into Standard Model particles, for which anti-nuclei created in reactions between primary cosmic ray protons and interstellar matter (pp, pA collisions) represent a source of background.

The main features of the theoretical frameworks employed for our study are briefly summarised in Sec.~\ref{sec:prod}, while we address the reader to~\cite{Bellini:2018epz} for the full details. Section~\ref{sec:results} presents the main results and conclusions follow.

\section{Modelling light (anti-)(hyper-)nuclei production} \label{sec:prod}

For our study, we consider nuclei and hyper-nuclei with mass number $A = 2, 3$ and 4, whose properties are summarised in Tab. \ref{tab:nucleusradii}. Those properties are the same as for their anti-matter counterparts and we assume that the same formation mechanisms are at play for matter and anti-matter\footnote{For brevity, in the following we refer to ``nuclei'' and ``hyper-nuclei'' but we imply both matter and anti-matter.}.
Nuclei and hyper-nuclei are special objects with respect to non-composite hadrons (pions, protons, etc.), because their size is comparable to a fraction or the whole system created in pp, \pPb~and \PbPb~collisions.
The size is typically defined in two ways: a) as the rms of the (charge) distribution ($\lambda_A$), typically measured in electron scattering experiments, or b) as the size parameter of the object wave-function ($r_A$), typically taken as the gaussian solution of an isotropic harmonic oscillator potential in coalescence calculations. 
For light nuclei, $\lambda_A\approx$~2~fm.  For the hyper-triton, theoretical calculations indicate a charge rms radius $\lambda_A\approx$~5~fm \cite{Nemura:1999qp}, driven by the average separation of the $\Lambda$ relative to the two other nucleons. 
Assuming a similar structure (e.g. a s-wave interaction for a bound state of a n or a $\Lambda$ with a
deuteron), the hypertriton results in a much larger object than the other non-strange nuclei with $A=3$.
A simple relation holds between $\lambda_A$ and $r_A$, see \cite{Bellini:2018epz}. 
In Tab.~\ref{tab:nucleusradii} the binding energy ($B_E$) is also reported. The most tightly bound nucleus is \hefour, whereas the most loosely bound object is \hthreelambda, that is also the largest one. For the latter, we report the separation energy of the $\Lambda$ baryon from the deuteron ($B_\Lambda=0.13$~MeV). 
The large size and the low binding energy of the \hthreelambda~with respect to the other (hyper-)nuclei has important consequences on its production, as discussed in what follows.

\begin{table}[htb]
\resizebox{\textwidth}{!}{%
\centering
\begin{tabularx}{1.\textwidth}{cccccccc}
\hline \hline \\[-2ex]
\makecell{Mass \\number } & Nucleus           &  \makecell{Compo-\\sition}               & $B_{E}$ (MeV)   & \makecell{Spin \\$J_{A}$} & \makecell{(Charge) \\rms radius \\ \rmsradius$^{meas}$~(fm)} &  \makecell{Harmonic \\ oscillator \\ size \\ parameter \\$r_{A}$ (fm) } & Refs. \\[1ex]  \hline \\[-2ex] 
      A = 2                     & d                                    & pn                                  &   2.224575 (9)     &     1   & 2.1413 $\pm$ 0.0025      &  3.2    &   \cite{VanDerLeun:1982bhg,Mohr:2015ccw}     \\[0.5ex]  \hline \\[-2ex]
\multirow{3}{*}{A = 3}  & \tritium 	                  & pnn                               &    8.4817986 (20) & 1/2   &  1.755  $\pm$ 0.086        &  2.15   &   \cite{Purcell:2015gtm}           \\
                                   & \hethree                         & ppn                                &   7.7180428  (23) & 1/2  & 1.959 $\pm$  0.030         &   2.48  &   \cite{Purcell:2015gtm} \\
                                   & \hthreelambda               & p$\Lambda$n                &    0.13 $\pm$ 0.05 & 1/2  &  4.9 --  10.0                    &  6.8 -- 14.1 & \cite{Davis:2005mb,Nemura:1999qp} \\[0.5ex]  \hline \\[-2ex]
\multirow{4}{*}{A = 4}  & \hefour                          & ppnn                              &    28.29566   (20)  &      0  &  1.6755 $\pm$ 0.0028  &  1.9  & \cite{1674-1137-41-3-030003,Angeli:2013epw} \\
                                   & \hfourlambda                & p$\Lambda$nn              &  2.04 $\pm$0.04   &   0   &    2.0 -- 3.8             & 2.4 -- 4.9  & \cite{Davis:2005mb,Nemura:1999qp} \\
                                   & \hfourtwolambda          &  p$\Lambda\Lambda$n &   0.39 -- 0.51         &    1     &    4.2 -- 7.1          & 5.5 -- 9.4  &   \cite{Nemura:1999qp} \\
                                   & \hefourlambda              & pp$\Lambda$n              &  2.39 $\pm$ 0.03  &    0   &    2.0 -- 3.8            & 2.4 -- 4.9  & \cite{Davis:2005mb,Nemura:1999qp}\\[0.5ex]  \hline \hline
\end{tabularx}
}
\caption{Properties of nuclei and hyper-nuclei with mass number $A \leq 4$. $B_{E}$ is the binding energy in MeV.   The size parameter $r_{A}$, is chosen to approximately reproduce the measured/expected rms,  \rmsradius$^{meas}$~(fm). The proton rms charge radius $\lambda_{p} = 0.879(8)$ fm is subtracted quadratically from the measured rms charge radius $\rmsradius^{meas}$ of the nucleus $\rmsradius = \sqrt{(\rmsradius^{meas})^2 - {\lambda_{p}^2}}$ to account for the finite extension of the constituents. Implicitly we assume here that $\lambda_{\Lambda}\approx \lambda_{n}\approx \lambda_{p}$.}
\label{tab:nucleusradii}
\end{table}

\subsection{The coalescence approach}
\begin{figure}[htb]
\begin{center}
\includegraphics[width=0.8\textwidth]{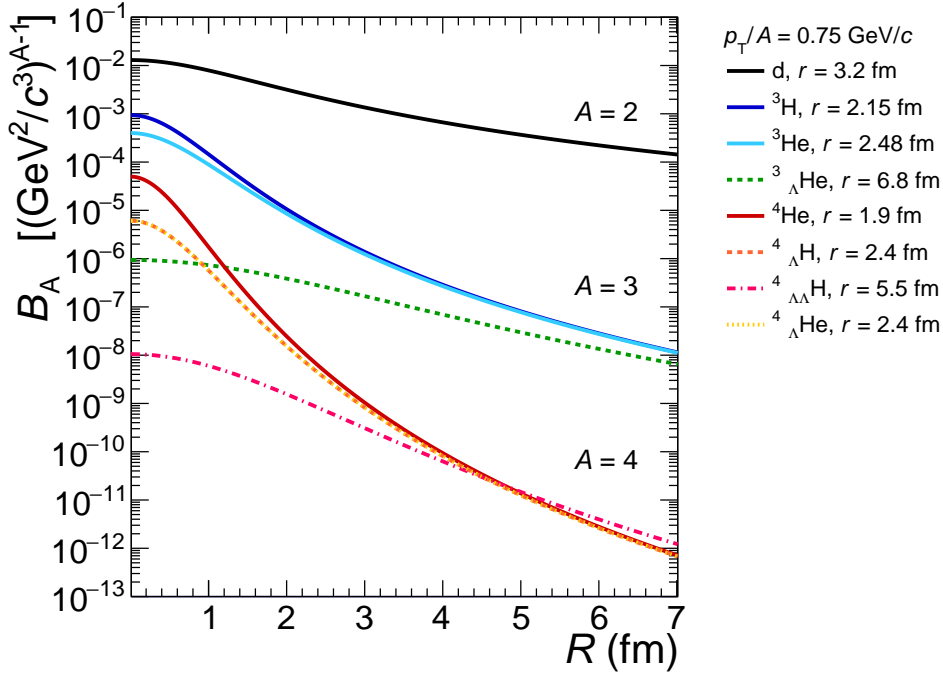}
\caption{(Color online) Coalescence parameter $B_A$ as a function of the source radius $R$ as predicted from the coalescence model (Eq.~\ref{eq:master}) for various composite objects with \pt$/A$~=~0.75~\GeVc. For each (hyper-)nucleus, the radius $r$ used for the calculation is reported in the legend.}
\label{Fig:BA}
\end{center}
\end{figure}

In the coalescence picture, nucleons produced in the collision coalesce into nuclei if they are close in space and have similar velocities \cite{Butler:1963, Kapusta:1980, Scheibl:1998tk}.
The coalescence probability is encoded in the coalescence parameter, $B_{A}$.
Considering that at LHC energies the number of produced protons and neutrons at midrapidity as well as their momentum distributions are expected to be equal, $B_{A}$ is defined as
\begin{equation}
E_{A}\frac{\mathrm{d}^{3}N_{A}}{\mathrm{d}p_{A}^{3}}=B_{A}{\left(E_{\mathrm{p,n}}\frac{\mathrm{d}^{3}N_{\mathrm{p,n}}}{\mathrm{d}p_{\mathrm{p,n}}^{3}}\right)^{A}}\left\vert_{\vec{p}_{\mathrm{p}}=\vec{p}_{\mathrm{n}}=\frac{\vec{p}_{A}}{A}} \right.,
\label{eq:BA}
\end{equation}
where $p_{\mathrm{p,n}}$ are the proton and neutron momenta and $E_{p,n}$ their energies. 
Equation~\ref{eq:BA} represents also the operative definition of $B_A$ that is used by experiments like ALICE to extract the coalescence probability starting from the measured nucleus and nucleon (proton) distributions.
Starting from the model described in~\cite{Scheibl:1998tk, Blum:2017qnn}, we have obtained in \cite{Bellini:2018epz} a generalised expression for $B_A$
\begin{equation}\label{eq:master}
B_{A} = {2J_{A} + 1 \over 2^{A}} {1 \over \sqrt{A}} {1 \over m_{T}^{A-1}} \Bigl({2\pi \over R^2 + ({r_A \over 2})^2 }\Bigr)^{\frac{3}{2}(A-1)} \;\;,
\end{equation}
\noindent which is a function of the spin of the particle $J_A$, its transverse mass $m_{\mathrm{T}}$, its size parameter $r_A$ and the source radius $R$. Very importantly, Eq.~\ref{eq:master} takes explicitly into account the source size ($R$), as the coalescence probability naturally decreases for nucleons with similar momenta that are produced far apart in configuration space. 
Moreover, the source size is identified with the  effective sub-volume of the whole system that is governed by the (momentum-dependent) homogeneity length of the interacting nucleons and experimentally accessible with Hanbury-Brown-Twiss (HBT) interferometry~\cite{Scheibl:1998tk, Blum:2017qnn}. 
Figure~\ref{Fig:BA} shows the source radius dependence of $B_A$ for nuclei and hyper-nuclei with $A = 2, 3$ and 4 whose properties are reported in Tab.~\ref{tab:nucleusradii}. 
For the cases in which more than one estimate for $r_A$ is available, we have adopted the lowest value for the calculations in Fig.~\ref{Fig:BA}.
We observe that the coalescence probability decreases with increasing mass number and $B_A$ decreases with increasing volume. 
For a given $A$, the larger the object radius, the lower is $B_A$, as clearly visible by comparing \hethree~and \hthreelambda. 
For objects with same $A$, mass and spin (e.g. the isobars \tritium~and \hethree), $B_A$ differs only due to the different radius $r_A$. 
This difference is more relevant in small systems, because in large systems the difference between nucleus radii is much smaller than the size of the source.  
Incidentally, this could be experimentally verified with high-precision measurements of the production of \tritium~relative to \hethree~in pp and \pPb~collisions. 
The production of objects with radius larger than the source is strongly suppressed, indicating that the process is driven by the length scale defined by the object radius relative to the source radius.

\subsection{Thermal + blast-wave model} 
Thermal-statistical models~\cite{Andronic:2010qu, Andronic:2017} have been successful in describing the production of light (anti-) (hyper-)nuclei across a wide range of energies in AA collisions, including production at the LHC. 
In this approach, particles are produced from a fireball in thermal equilibrium with temperatures of $T_{chem} \approx$ 156 MeV. 
Particle yields are derived from the partition function assuming a Grand Canonical ensemble\footnote{Extensions to small systems employ a canonical ensemble partition function to account for the exact conservation of quantum numbers in a finite size system, see for instance~\cite{Vovchenko:2018fiy}.} and they depend only on the mass of the particle and the temperature of chemical freeze-out, d$N$/d$y\propto$~(2$J_A$+1)exp$(-m_{A}/T_{\mathrm{chem}})$.
The thermal model cannot -- alone -- be compared to the \pt-dependent coalescence description because it provides only predictions for \pt-integrated yields. 
A thermal particle production implies a Boltzmann distribution of the momenta only for a static source, which is not the case of the rapidly expanding system produced in heavy-ion collisions. 
The thermal model (i.e. the GSI-Heidelberg implementation we have considered here) needs to be complemented by a hydrodynamic description of the rapidly expanding source. 
To that end, we use the Blast-Wave model~\cite{Schnedermann:1993ws}, which has been proven to describe reasonably well the measured momentum distributions of protons~\cite{Abelev:2013vea}. 
Our ``thermal+blast-wave'' approach results from the combination of the two models: the \pt-dependence is modelled by the blast-wave whereas the normalisation (i.e. the \pt-integrated yield) is taken from the thermal model predictions.
In particular, to describe the (hyper-)nucleus transverse momentum distributions we use the blast-wave parameters obtained from the simultaneous fit to pion, kaon and proton spectra measured in \PbPb~collisions by ALICE for several centralities/multiplicities and reported in~\cite{Abelev:2013vea}. We therefore inherit the multiplicity-dependence of the radial expansion of the system from the measurements. The object size does not enter in the formulation of the blast-wave model. 
In our thermal+blast-wave model, for each centrality/multiplicity class, the coalescence parameter is extracted according to Eq.~\ref{eq:BA} (as done for data) from the predicted (hyper-)nucleus spectrum and the proton spectrum measured by ALICE~\cite{Abelev:2013vea}.

\section{Results and discussion} \label{sec:results}
\begin{figure}[!htb]
\begin{center}
\includegraphics[width=0.88\textwidth]{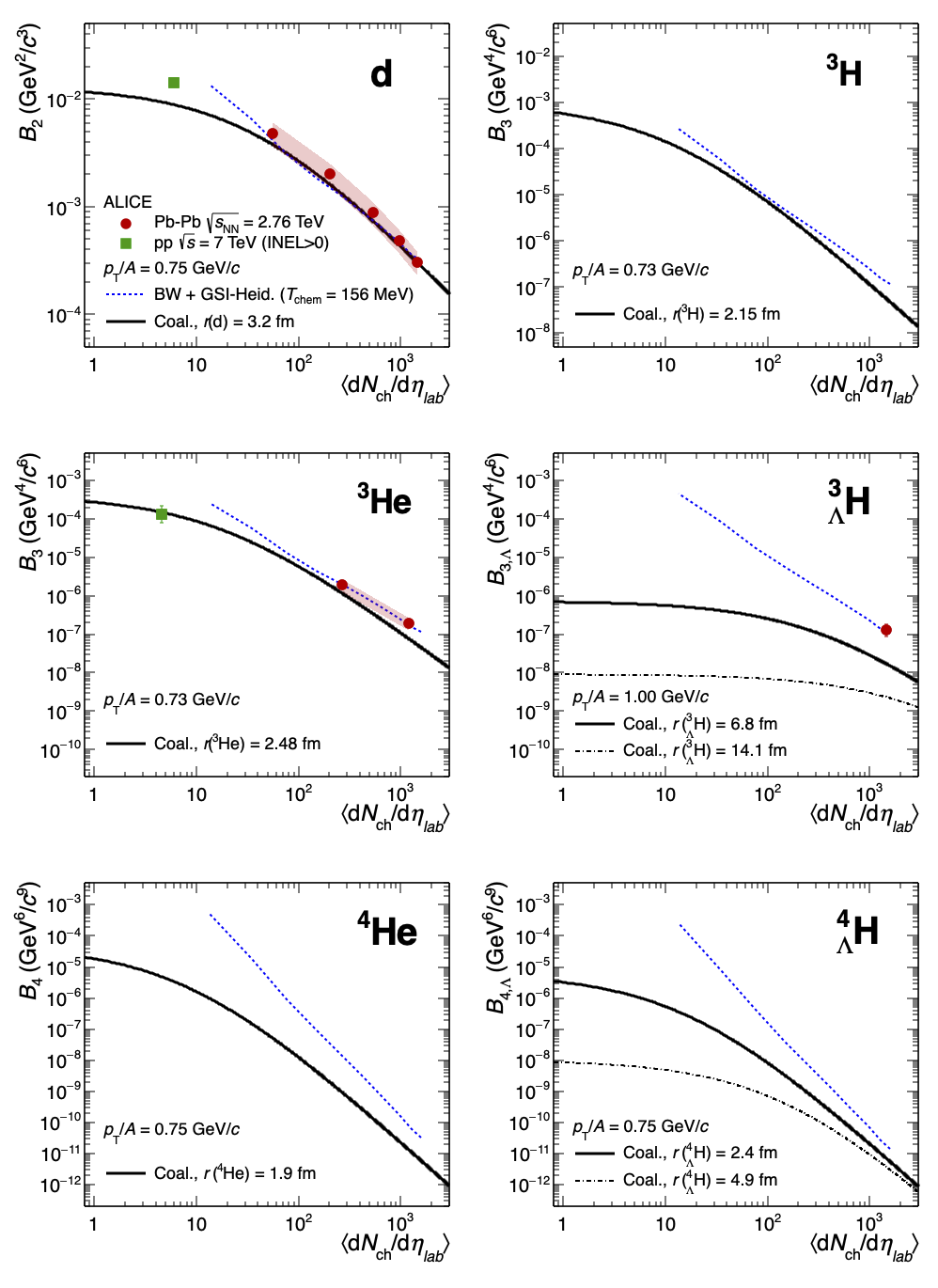}
\caption{(Color online) Coalescence parameter $B_A$ as a function of the average charged particle multiplicity density for various (hyper-)nuclei, up to $A = 4$. The coalescence calculations (continuous or dashed-dotted black lines) are compared to the thermal+blast-wave predictions (dashed blue line), as well as to pp (green square) and \PbPb~(red circles) collision data from ALICE~\cite{ALICE:nucleipp2017,ALICE:deuteronppPbPb2015,Adam:2015yta}. }
\label{Fig:comparison}
\end{center}
\end{figure}

As discussed in the previous sections, the coalescence model provides an analytical expression for $B_A$ as a function of the source size $R$. 
The thermal+blast-wave approach allows us to extract the $B_A$ as a function of the average charged particle multiplicity density, \avdNdeta. Measurements are performed differentially in centrality/multiplicity.
In order to compare models and data, we therefore need to map multiplicity into source radius or viceversa. 
As discussed more extensively in~\cite{Bellini:2018epz}, we perform this mapping based on the parameterisation $R = 0.473 \avdNdeta^{1/3}$ for \pt/$A$~=~0.75 \GeVc. The slope is smaller for larger \pt's and larger for lower \pt.
The value of the empirical slope parameter is obtained by tuning the parameterisation such that the measured (anti-)deuteron \btwo~in the most central \PbPb~class falls onto the coalescence prediction. In other words, we constrain the coalescence volume with the more differential (anti-)deuteron data and assume that it is the same for all (hyper-)nuclei. 
In~\cite{Bellini:2018epz}, we used the parameterisation to map the \avdNdeta~into $R$. Here, we chose to apply the inverse relation and compare $B_A$ from models with ALICE data as a function of the experimentally accessible observable~\avdNdeta. The two choices are equivalent. This comparison is reported in Fig.~\ref{Fig:comparison}. 
With respect to~\cite{Bellini:2018epz}, we have now extended our study up to $A=4$ \mbox{(hyper-)nuclei}. 

We observe that for deuterons, the coalescence and thermal+blast-wave approaches lead to similar predictions and describe reasonably the \PbPb~data. 
For \hethree, the models exhibit a qualitatively similar mul\-ti\-pli\-ci\-ty-dependence with the magnitude of $B_A$ differing by a factor 1.5 to 2 going from large to small systems. The currently available data for \hethree~are consistent with both models within 2$\sigma$ to 3$\sigma$, where $\sigma$ is the total uncertainty on data. 
Both approaches show large differences (a factor 5 to 6 for central Pb--Pb collisions and a factor larger than 50 for \avdNdeta~<~100) for the \hthreelambda\ caused by the significantly larger size of \hthreelambda\ with respect to \hethree. 
The only data point available so far in \PbPb~collisions is in agreement with the thermal+blast-wave model but differs by 6$\sigma$ from our coalescence calculation. 
Also for \hefour~and \hefourlambda~the coalescence probability is a steeply falling function of multiplicity in both approaches, which differ in magnitude from few units at high multiplicities to O(10$^{2}$) in small systems.
We conclude that the difference between the two approaches increases from large to small systems, which advocates for new multiplicity-differential data in all collision systems to distinguish between the two production scenarios. 
For \hthreelambda~and \hefourlambda~we also report in Fig.~\ref{Fig:comparison} the coalescence calculations obtained assuming wider wave-functions, which result in even lower production probabilities. 

\begin{figure}[!htb]
\begin{center}
\includegraphics[width=0.8\textwidth]{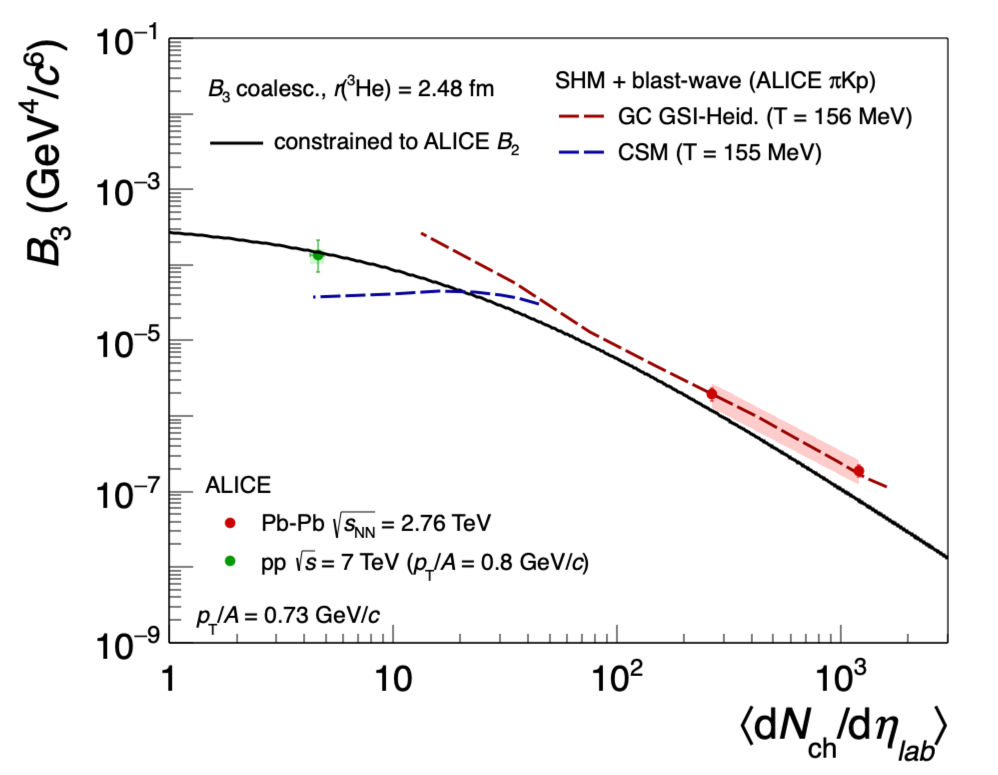}
\caption{(Color online) Coalescence parameter $B_3$ for \hethree~as a function of the average charged particle multiplicity density~\avdNdeta. The coalescence calculation (continuous black line) is compared to two thermal+blast-wave predictions (dashed lines), obtained by using the Grand Canonical (GC, red)~\cite{Andronic:2017} and Canonical Statistical Model (CSM, blue) \cite{Vovchenko:2018fiy} expectations for the \hethree~yield, respectively. ALICE data from pp (green circles) and \PbPb~(red circles) collisions~\cite{ALICE:nucleipp2017,ALICE:deuteronppPbPb2015} are reported. }
\label{Fig:3He}
\end{center}
\end{figure}

The Grand Canonical thermal ansatz is no longer valid when going to small collision systems, because one needs to account for the exact conservation of quantum numbers in a finite volume and thus a canonical ensemble has to be used.
As an example, we have also computed $B_A$ for \hethree~according to the thermal+blast-wave approach for the \pPb~system, by using yield predictions from the Canonical Statistical Model from~\cite{Vovchenko:2018fiy} and the blast-wave parameters from~\cite{Abelev:2013haa}. The result is reported in Fig. \ref{Fig:3He} with the blue dashed line. 
We notice that the thermal+blast-wave prediction in this case does not exhibit a significant dependence on multiplicity and is relatively close to the coalescence curve. As expected, in high multiplicity \pPb~collisions the CSM+blast-wave curve tends to merge with the Grand Canonical prediction. 

\section{Conclusions}
We have presented a direct comparison of the coalescence and ther\-mal+blast-wa\-ve production scenarios for light nuclei and hyper-nuclei as a function of multiplicity, used as a proxy for the source size. 
To clarify the production mechanism of composite loosely-bound QCD objects, we show the importance of new and precise measurements of (hyper-)nuclei production up to $A=4$ as a function of multiplicity in different collision systems, in order to fully exploit the sensitivity of production mechanisms to the size of the object relative to the size of the source.
Particularly sensitive to production mechanisms is the hyper-triton with its large size. 
The increase in integrated luminosity foreseen in all collision systems during the LHC Runs 3 and 4 in years 2021-2029 should allow the ALICE Collaboration to perform these multi-differential measurements, which will provide crucial insights on the ``anti-nuclei puzzle'' and the (hyper-)nuclei formation.

\bibliographystyle{utphysnt} 	
\bibliography{biblio}

\providecommand{\href}[2]{#2}\begingroup\raggedright\begin{thebibliography}{10}

\bibitem{Butler:1963}
S.~T. Butler and C.~A. Pearson
\href{http://dx.doi.org/10.1103/PhysRev.129.836}{{\em Phys. Rev.} {\bfseries
  129} (1963) 836--842}.

\bibitem{Kapusta:1980}
J.~I. Kapusta
\href{http://dx.doi.org/10.1103/PhysRevC.21.1301}{{\em Phys. Rev.} {\bfseries
  C21} (1980) 1301--1310}.

\bibitem{Scheibl:1998tk}
R.~Scheibl and U.~W. Heinz
  \href{http://dx.doi.org/10.1103/PhysRevC.59.1585}{{\em Phys. Rev.} {\bfseries
  C59} (1999) 1585--1602},
\href{http://arxiv.org/abs/nucl-th/9809092}{{\ttfamily arXiv:nucl-th/9809092
  [nucl-th]}}.

\bibitem{Blum:2017qnn}
K.~Blum, K.~C.~Y. Ng, R.~Sato, and M.~Takimoto
  \href{http://dx.doi.org/10.1103/PhysRevD.96.103021}{{\em Phys. Rev.}
  {\bfseries D96} no.~10, (2017) 103021},
\href{http://arxiv.org/abs/1704.05431}{{\ttfamily arXiv:1704.05431
  [astro-ph.HE]}}.

\bibitem{Andronic:2010qu}
A.~Andronic, P.~Braun-Munzinger, J.~Stachel, and H.~Stocker
  \href{http://dx.doi.org/https://doi.org/10.1016/j.physletb.2011.01.053}{{\em
  Physics Letters B} {\bfseries 697} no.~3, (2011) 203 -- 207}.

\bibitem{Andronic:2017}
A.~Andronic, P.~Braun-Munzinger, K.~Redlich, and J.~Stachel {\em Nature}
  {\bfseries 561} no.~7723, (2018) 321--330,
  \href{http://arxiv.org/abs/1710.09425}{{\ttfamily arXiv:1710.09425
  [nucl-th]}}.

\bibitem{ALICE:deuteronppPbPb2015}
{\bfseries ALICE} Collaboration, J.~Adam {\em et~al.}
  \href{http://dx.doi.org/10.1103/PhysRevC.93.024917}{{\em Phys. Rev.}
  {\bfseries C93} no.~2, (2016) 024917},
\href{http://arxiv.org/abs/1506.08951}{{\ttfamily arXiv:1506.08951 [nucl-ex]}}.

\bibitem{Adam:2015yta}
{\bfseries ALICE} Collaboration, J.~Adam {\em et~al.}
  \href{http://dx.doi.org/10.1016/j.physletb.2016.01.040}{{\em Phys. Lett.}
  {\bfseries B754} (2016) 360--372},
\href{http://arxiv.org/abs/1506.08453}{{\ttfamily arXiv:1506.08453 [nucl-ex]}}.

\bibitem{ALICE:nucleipp2017}
{\bfseries ALICE} Collaboration, S.~Acharya {\em et~al.}
  \href{http://dx.doi.org/10.1103/PhysRevC.97.024615}{{\em Phys. Rev.}
  {\bfseries C97} no.~2, (2018) 024615},
\href{http://arxiv.org/abs/1709.08522}{{\ttfamily arXiv:1709.08522 [nucl-ex]}}.

\bibitem{Acharya:2017dmc}
{\bfseries ALICE} Collaboration, S.~Acharya {\em et~al.}
  \href{http://dx.doi.org/10.1140/epjc/s10052-017-5222-x}{{\em Eur. Phys. J.}
  {\bfseries C77} no.~10, (2017) 658},
\href{http://arxiv.org/abs/1707.07304}{{\ttfamily arXiv:1707.07304 [nucl-ex]}}.

\bibitem{Puccio:2019oyd}
{\bfseries ALICE} Collaboration, M.~Puccio
\href{http://dx.doi.org/10.1016/j.nuclphysa.2018.10.043}{{\em Nucl. Phys.}
  {\bfseries A982} (2019) 447--450}.

\bibitem{Acharya:2019rgc}
{\bfseries ALICE} Collaboration, S.~Acharya {\em et~al.}
\href{http://arxiv.org/abs/1902.09290}{{\ttfamily arXiv:1902.09290 [nucl-ex]}}.

\bibitem{Mrowczynski:2016xqm}
S.~Mrowczynski \href{http://dx.doi.org/10.5506/APhysPolB.48.707}{{\em Acta
  Phys. Polon.} {\bfseries B48} (2017) 707},
\href{http://arxiv.org/abs/1607.02267}{{\ttfamily arXiv:1607.02267 [nucl-th]}}.

\bibitem{Bellini:2018epz}
F.~Bellini and A.~P. Kalweit
\href{http://arxiv.org/abs/1807.05894}{{\ttfamily arXiv:1807.05894 [hep-ph]}}.

\bibitem{Bazak:2018hgl}
S.~Bazak and S.~Mrowczynski
  \href{http://dx.doi.org/10.1142/S0217732318501420}{{\em Mod. Phys. Lett.}
  {\bfseries A33} no.~25, (2018) 1850142},
\href{http://arxiv.org/abs/1802.08212}{{\ttfamily arXiv:1802.08212 [nucl-th]}}.

\bibitem{Zhao:2018lyf}
W.~Zhao, L.~Zhu, H.~Zheng, C.~M. Ko, and H.~Song
  \href{http://dx.doi.org/10.1103/PhysRevC.98.054905}{{\em Phys. Rev.}
  {\bfseries C98} no.~5, (2018) 054905},
\href{http://arxiv.org/abs/1807.02813}{{\ttfamily arXiv:1807.02813 [nucl-th]}}.

\bibitem{Sun:2018mqq}
K.-J. Sun, C.~M. Ko, and B.~Doenigus
\href{http://arxiv.org/abs/1812.05175}{{\ttfamily arXiv:1812.05175 [nucl-th]}}.

\bibitem{Xu:2018jff}
X.~Xu and R.~Rapp
\href{http://arxiv.org/abs/1809.04024}{{\ttfamily arXiv:1809.04024 [nucl-th]}}.

\bibitem{Oliinychenko:2018ugs}
D.~Oliinychenko, L.-G. Pang, H.~Elfner, and V.~Koch
\href{http://arxiv.org/abs/1809.03071}{{\ttfamily arXiv:1809.03071 [hep-ph]}}.

\bibitem{Garcilazo:1982yc}
H.~Garcilazo
\href{http://dx.doi.org/10.1103/PhysRevLett.48.577}{{\em Phys. Rev. Lett.}
  {\bfseries 48} (1982) 577--580}.

\bibitem{Bass:1998ca}
S.~A. Bass {\em et~al.}
  \href{http://dx.doi.org/10.1016/S0146-6410(98)00058-1}{{\em Prog. Part. Nucl.
  Phys.} {\bfseries 41} (1998) 255--369},
\href{http://arxiv.org/abs/nucl-th/9803035}{{\ttfamily arXiv:nucl-th/9803035
  [nucl-th]}}.

\bibitem{Schukraft:2017nbn}
J.~Schukraft \href{http://dx.doi.org/10.1016/j.nuclphysa.2017.05.036}{{\em
  Nucl. Phys.} {\bfseries A967} (2017) 1--10},
\href{http://arxiv.org/abs/1705.02646}{{\ttfamily arXiv:1705.02646 [hep-ex]}}.

\bibitem{Citron:2018lsq}
Z.~Citron {\em et~al.} in {\em {HL/HE-LHC Workshop: Workshop on the Physics of
  HL-LHC, and Perspectives at HE-LHC Geneva, Switzerland, June 18-20, 2018}}.
\newblock 2018.
\newblock
\href{http://arxiv.org/abs/1812.06772}{{\ttfamily arXiv:1812.06772 [hep-ph]}}.
\newblock

\bibitem{Aramaki:2015pii}
T.~Aramaki {\em et~al.}
  \href{http://dx.doi.org/10.1016/j.physrep.2016.01.002}{{\em Phys. Rept.}
  {\bfseries 618} (2016) 1--37},
\href{http://arxiv.org/abs/1505.07785}{{\ttfamily arXiv:1505.07785 [hep-ph]}}.

\bibitem{Cirelli:2014qia}
M.~Cirelli, N.~Fornengo, M.~Taoso, and A.~Vittino
  \href{http://dx.doi.org/10.1007/JHEP08(2014)009}{{\em JHEP} {\bfseries 08}
  (2014) 009},
\href{http://arxiv.org/abs/1401.4017}{{\ttfamily arXiv:1401.4017 [hep-ph]}}.

\bibitem{Korsmeier:2017xzj}
M.~Korsmeier, F.~Donato, and N.~Fornengo
  \href{http://dx.doi.org/10.1103/PhysRevD.97.103011}{{\em Phys. Rev.}
  {\bfseries D97} no.~10, (2018) 103011},
\href{http://arxiv.org/abs/1711.08465}{{\ttfamily arXiv:1711.08465
  [astro-ph.HE]}}.

\bibitem{Nemura:1999qp}
H.~Nemura, Y.~Suzuki, Y.~Fujiwara, and C.~Nakamoto
  \href{http://dx.doi.org/10.1143/PTP.103.929}{{\em Prog. Theor. Phys.}
  {\bfseries 103} (2000) 929--958},
\href{http://arxiv.org/abs/nucl-th/9912065}{{\ttfamily arXiv:nucl-th/9912065
  [nucl-th]}}.

\bibitem{VanDerLeun:1982bhg}
C.~Van Der~Leun and C.~Alderliesten
\href{http://dx.doi.org/10.1016/0375-9474(82)90105-1}{{\em Nucl. Phys.}
  {\bfseries A380} (1982) 261--269}.

\bibitem{Mohr:2015ccw}
P.~J. Mohr, D.~B. Newell, and B.~N. Taylor
  \href{http://dx.doi.org/10.1103/RevModPhys.88.035009}{{\em Rev. Mod. Phys.}
  {\bfseries 88} no.~3, (2016) 035009},
\href{http://arxiv.org/abs/1507.07956}{{\ttfamily arXiv:1507.07956
  [physics.atom-ph]}}.

\bibitem{Purcell:2015gtm}
J.~E. Purcell and C.~G. Sheu
\href{http://dx.doi.org/10.1016/j.nds.2015.11.001}{{\em Nucl. Data Sheets}
  {\bfseries 130} (2015) 1--20}.

\bibitem{Davis:2005mb}
D.~H. Davis
\href{http://dx.doi.org/10.1016/j.nuclphysa.2005.01.002}{{\em Nucl. Phys.}
  {\bfseries A754} (2005) 3--13}.

\bibitem{1674-1137-41-3-030003}
M.~Wang, G.~Audi, F.~Kondev, W.~Huang, S.~Naimi, and X.~Xu {\em Chinese Physics
  C} {\bfseries 41} no.~3, (2017) 030003.

\bibitem{Angeli:2013epw}
I.~Angeli and K.~P. Marinova
\href{http://dx.doi.org/10.1016/j.adt.2011.12.006}{{\em Atom. Data Nucl. Data
  Tabl.} {\bfseries 99} no.~1, (2013) 69--95}.

\bibitem{Vovchenko:2018fiy}
V.~Vovchenko, B.~Doenigus, and H.~Stoecker
  \href{http://dx.doi.org/10.1016/j.physletb.2018.08.041}{{\em Phys. Lett.}
  {\bfseries B785} (2018) 171--174},
\href{http://arxiv.org/abs/1808.05245}{{\ttfamily arXiv:1808.05245 [hep-ph]}}.

\bibitem{Schnedermann:1993ws}
E.~Schnedermann, J.~Sollfrank, and U.~W. Heinz
  \href{http://dx.doi.org/10.1103/PhysRevC.48.2462}{{\em Phys. Rev.} {\bfseries
  C48} (1993) 2462--2475},
\href{http://arxiv.org/abs/nucl-th/9307020}{{\ttfamily arXiv:nucl-th/9307020
  [nucl-th]}}.

\bibitem{Abelev:2013vea}
{\bfseries ALICE} Collaboration, B.~Abelev {\em et~al.}
  \href{http://dx.doi.org/10.1103/PhysRevC.88.044910}{{\em Phys. Rev.}
  {\bfseries C88} (2013) 044910},
\href{http://arxiv.org/abs/1303.0737}{{\ttfamily arXiv:1303.0737 [hep-ex]}}.

\bibitem{Abelev:2013haa}
{\bfseries ALICE} Collaboration, B.~B. Abelev {\em et~al.}
  \href{http://dx.doi.org/10.1016/j.physletb.2013.11.020}{{\em Phys. Lett.}
  {\bfseries B728} (2014) 25--38},
\href{http://arxiv.org/abs/1307.6796}{{\ttfamily arXiv:1307.6796 [nucl-ex]}}.

\end{thebibliography}\endgroup
\end{document}